\documentclass[runningheads]{llncs}
\usepackage{graphicx}
\usepackage{amsmath}

\usepackage{stfloats}

\usepackage{amsfonts}
\usepackage{graphicx}
\usepackage{booktabs}
\usepackage{float}

\usepackage{multirow}

\begin{document}
	%

	
		\title{Brain Tumor Synthetic Segmentation in 3D Multimodal MRI Scans}

	%
	\titlerunning{Brain Tumor Synthetic Segmentation in 3D  Multimodal MRI Scans}
	%
	\author{Mohammad~Hamghalam\inst{1,2}\orcidID{0000-0003-2543-0712} \and
	Baiying~Lei\inst{1} \and
	Tianfu~Wang\inst{1}}
	\authorrunning{M. Hamghalam et al.}
	%
	\institute{National-Regional Key Technology Engineering Laboratory for Medical Ultrasound, Guangdong Key Laboratory for Biomedical Measurements and Ultrasound Imaging, School of Biomedical Engineering, Health Science Center, Shenzhen University, Shenzhen, China, 518060. \and
	Faculty of Electrical, Biomedical and Mechatronics Engineering, Qazvin Branch, Islamic Azad
	University, Qazvin, Iran.\\
	\email{m.hamghalam@gmail.com}\\
	\email{\{leiby,tfwang\}@szu.edu.cn}}

	\maketitle              
	\begin{abstract}
	The magnetic resonance (MR) analysis of brain tumors is widely used for diagnosis and examination of tumor subregions. The overlapping area among the intensity distribution of healthy, enhancing, non-enhancing, and edema regions makes the automatic segmentation a challenging task. Here, we show that a convolutional neural network trained on high-contrast images can transform the intensity distribution of brain lesions in its internal subregions. Specifically, a generative adversarial network (GAN) is extended to synthesize high-contrast images.
	A comparison of these synthetic images and real images of brain tumor tissue in MR scans showed significant segmentation improvement and decreased the number of real channels for segmentation.
	The synthetic images are used as a substitute for real channels and can bypass real modalities in the multimodal brain tumor segmentation framework. Segmentation results on BraTS 2019 dataset demonstrate that our proposed approach can efficiently segment the tumor areas. In the end, we predict patient survival time based on volumetric features of the tumor subregions as well as the age of each case through several regression models.
	\end{abstract}
	\section{Introduction}    
	\label{sec:intro}
	
Glioma is the most aggressive and widespread tumor is grouped into low-grade gliomas (LGGs) and high-grade gliomas (HGGs). Multimodal MR channels in BraTS 2019 datasets \cite{Bakas2018_Identifying,Bakas2017_Radiomic_2,Bakas2017_Radiomic_1,bakas2017advancing,Menze2015Brain}, included of FLAIR, T1, T1c, and T2, are routinely used to segment internal parts of the tumor, i.e., whole tumor (WT), tumor core (TC), and enhancing tumor (ET). Several segmentation approaches have been proposed to segment regions of interest through classic \cite{HamghalamAutomatic2009,Hamghalam2009,Soleymanifard2019,Soleimany2017} and modern machine learning methods, especially brain tumor segmentation techniques \cite{Hatami2019,Najrabi2018}.

	\begin{figure*}[ht]
	\includegraphics[width=12cm]{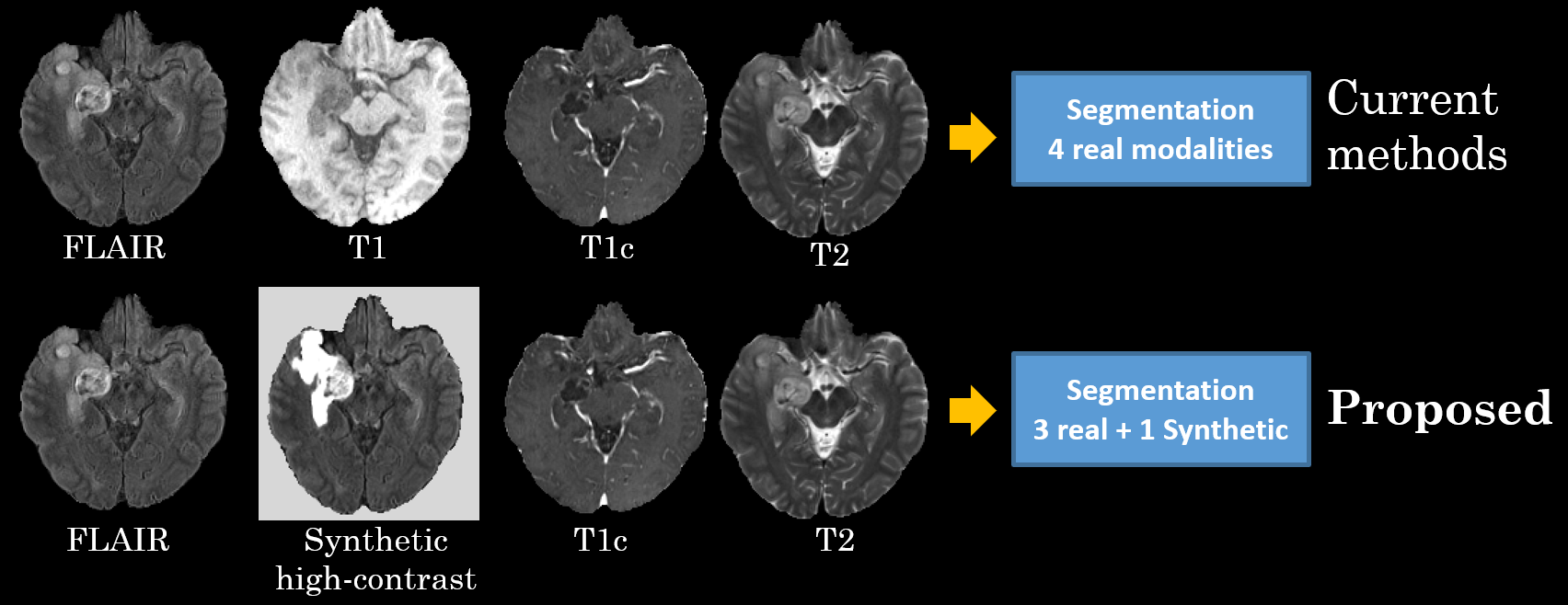}
	
	\caption{The pipeline outlines the steps in the current (top) and proposed synthetic (bottom) segmentation techniques. 
		We displace the real T1 channels with the synthetic image.}
	\label{fig:pipeline}
\end{figure*}

	\begin{figure*}[ht]

		\includegraphics[width=12cm]{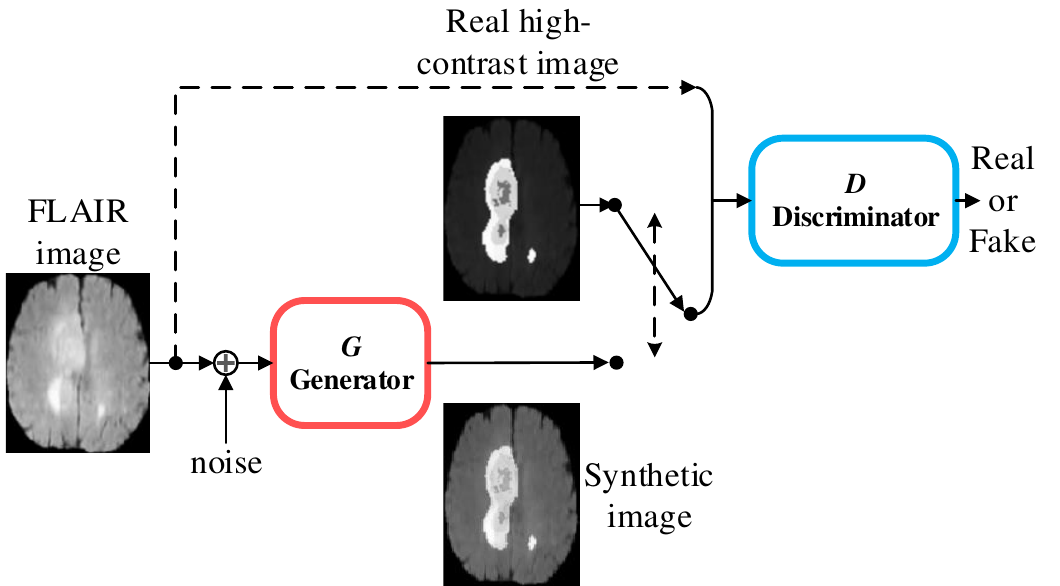}
		
		\caption{Deep-learning-based high-contrast synthesis using FLAIR images. After training by GAN, the model outputs the synthetic high tissue contrast images with an inference time of around 20 ms.}
		\label{fig:framework}
		
	\end{figure*}

The focus of current research is to form a generator that increases the contrast within subregions of the brain tissue. 
The generator, which is a deep neural network model, employes a real channel as input to produce the synthetic one. Our framework comprises two stages: (1) we generate high tissue contrast images based on FLAIR sequence in our convolutional neural network (CNN) model, (2) we train a 3D fully convolutional network (FCN) \cite{Chen2018,Harley2017,roy2017error,Long2015} based on the synthetic images to segment region of interests.
	\section{Method}
	\label{sec:Method}
Our goal is to segment tumor subregions based on multimodal 3D magnetic resonance (MR) volumes. Fig.\ref{fig:pipeline} demonstrates an overview of the proposed method based on synthetic high-contrast images. In contrast to the current methods, we use both real and synthetic volumes for the segmentation task. Following, we first introduce the synthetic image generator module, based on the generative adversarial networks (GANs) model \cite{goodfellow2014generative}, and then 3D FCN architecture for segmentation is discussed.

	\subsection{Synthetic Image Generator}
	\label{sssec:3D-to-2D}
	
We extend the image-to-image translation method \cite{isola2017image} to deal with the synthesis of high-contrast 2D images. Our model trains on high-contrast images, building based on manual labels, in an adversarial framework. The synthesis model contains a Generator, based on the 2D-U-Net \cite{ronneberger2015u}, and a Discriminator, build on 2D FCN  network. Fig. \label{fig:framework} illustrates the image translation framework, where both the generator and the discriminator blocks are trained on FLAIR with a patch size of 128 $\times$ 128 pixels. In implementation details, we follow \cite{isola2017image}, including the number of epochs, the number of layers, and the kernel sizes. For each subject in the BraTS'19 dataset, we provide a 3D synthetic volume for the next stage, segmentation. 
	
	\subsection{Synthetic Segmentation}
	\label{sssec:Classifier block}
	The output volumes from synthetic image generator block are concatenated with real modalities (FLAIR, T1c, and T2) and fed into segmentation block to predict region of interests.
	The segmentation network allows jointly capturing features from FLAIR, synthetic, T1c, and T2 modality. For the 3D segmentation block, we rely on ensembling the 3D FCN on axial, sagittal, and coronal planes.

	\section{Experimental Results}
	\label{sec:pagestyle}
	
	\subsection{Implementation Details}
	\label{sec:Implementation}
	We implement the proposed design employing the KERAS with 12GB NVIDIA TITAN X GPU. We have scaled image patches to sizes 128 $\times$ 128 pixels for translation.
	The model is trained through the ADADELTA \cite{Matthew2012} optimizer (learning rate = 0.9, $\rho=0.90$, epsilon=1e-5). Dropout is employed to avoid over-fitting over the training ($p_{drop}=0.4$).
	\subsection{Datasets}
	\label{sec:Dataset}
	The performance of the proposed method is evaluated on the BraTS'19 dataset, which has two datasets of pre-operative MRI sequences: Training (335 cases) and Validation (125 cases). Each patient is giving $155\times240\times240$ with four channels: T1, T2, T1c, and FLAIR. In the manual label of BraTS'19, there are three tumor regions: non-enhancing tumor, enhancing tumor, and edema. The evaluation is figured out by CBICA IPP\footnote{https://ipp.cbica.upenn.edu} online platforms.
	Metrics computed by the online evaluation platforms in BraTS'19 are Dice Similarity Coefficient (DSC) and the 95th percentile of the Hausdorff Distance (HD95). DSC is considered to measure the union of prediction and manual segmentation. It is measured as $DSC = \frac{2TP}{FP+2TP+FN}$ where TP, FP, and FN are the numbers of true positive, false positive, and false negative detections, respectively.

\subsection{Segmentation Results on BRATS'19}	

Fig. \ref{fig:framework} shows examples of brain tumor prediction in LGG and HGG slides on BraTS19 along with corresponding labels, where the subject IDs are	"BraTS19-TCIA10-175-1" and "BraTS19-CBICA-APK-1" for LGG and HGG, respectively.	
The results in Table \ref{tab:segmentation_result_val} show that our method performed competitive performance on validation set (125 cases) of BraTS dataset. Results are reported in the online processing platform by BraTS'19 organizer. Moreover, Table \ref{tab:segmentation_result_training} reports the average results on 335 training case of the BraTS'19.  

	\begin{figure*}[!t]
	
	\centering
	\includegraphics[width=11cm]{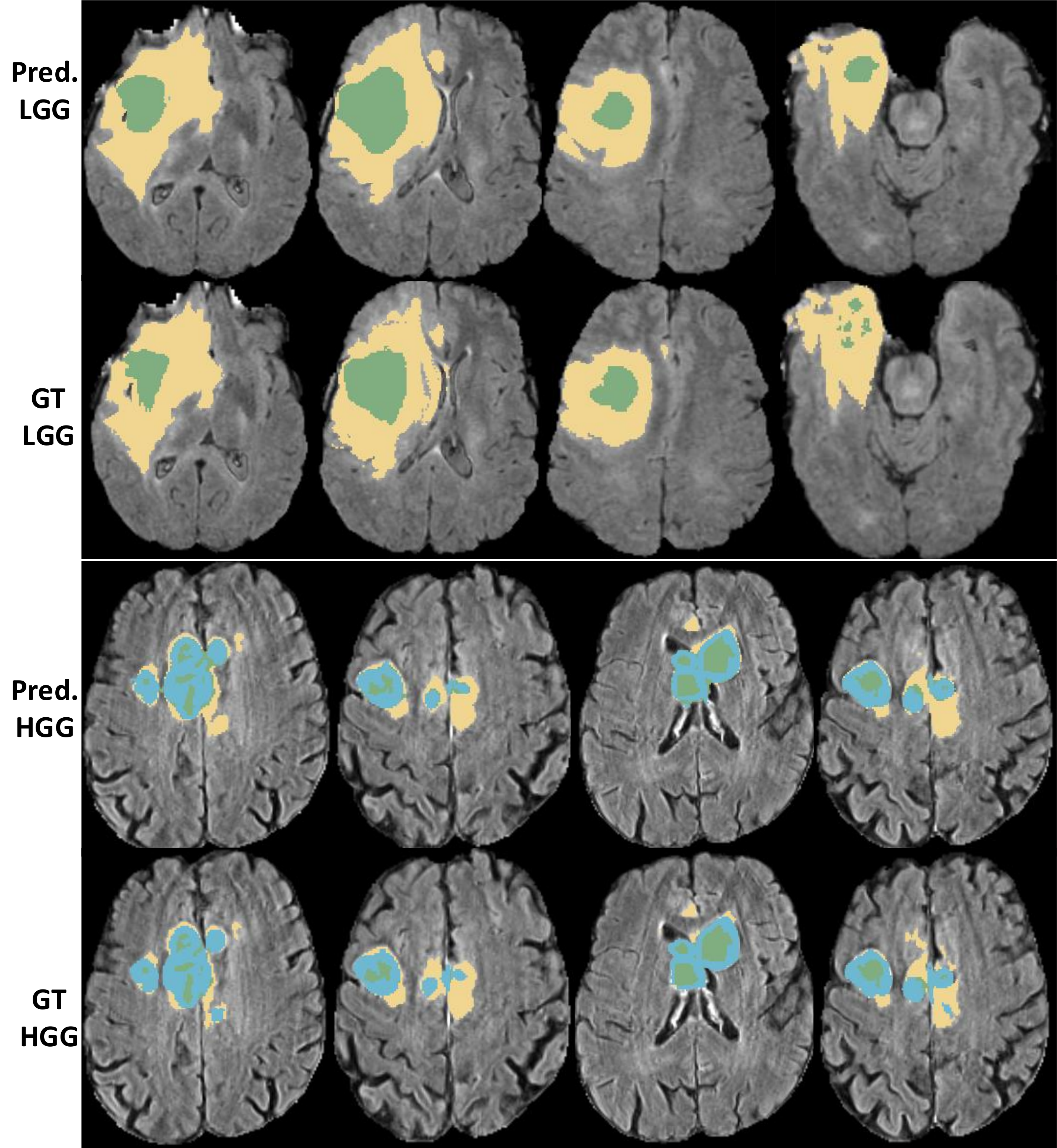}
	
	\caption{Segmentation results are overlaid on FLAIR axial slices on BraTS'19 Training Data. The yellow label is edema, blue color means enhancing tumor, and the green one shows the necrotic and non-enhancing tumor core. 
		The first and second rows illustrate LGG brain tumor, prediction (Pred.), and ground truth (GT), respectively. The third and fourth rows are related to HGG tumors.
		Computed DSCs by the Challenge organizer are reported for the LGG subject as: WT = 96.55\% and ET\% = 88.85, as well as HGG subject as: TC = 93.80\%, WT = 93.97\%, and ET = 95.00\%.	
}
	\label{fig:framework}
	
\end{figure*}

	\begin{table}[!t]
		\centering
		\caption{DSCs and HD95 of the synthetic segmentation method on BraTS'19 Validation set (training on 335 cases of BraTS'19 training set).} 
		\label{tab:segmentation_result_val}

	\begin{tabular}{@{\extracolsep{\fill}}c|cccc|cccc|cccc|ccc}
		\toprule[1pt] 
		\rule[-1ex]{0pt}{2.5ex}
		\multirow{3}{*}{\textbf{}} & \multicolumn{3}{c}{\textbf{Dice}} && \multicolumn{3}{c}{\textbf{Sensitivity}  }  && \multicolumn{3}{c}{\textbf{Specificity}  } && \multicolumn{3}{c}{\textbf{HD95 (mm)}  } \\
		\cmidrule[\heavyrulewidth]{2-4} 
		\cmidrule[\heavyrulewidth]{6-8}
		\cmidrule[\heavyrulewidth]{10-12}
		\cmidrule[\heavyrulewidth]{14-16}
		&\textbf{ET} &\textbf{WT} & \textbf{TC} && \textbf{ET} &\textbf{WT} & \textbf{TC}&& \textbf{ET} &\textbf{WT} & \textbf{TC}&& \textbf{ET} &\textbf{WT} & \textbf{TC}\\
		
		\midrule 
		Mean                 & 76.65 &    89.65    & 79.01 &&    76.88    & 91.32 &    77.71 &&    99.85 &    99.39 &    99.76    &&    4.6    & 6.9&    8.4\\
		\midrule 
		Std.      & 25.86    & 9.44    & 23.31    && 25.35    & 8.84 & 26.13    && 0.23    & 0.69    & 0.33    &&        7.2    & 13.8    & 12.4\\
		\midrule 
		Median                & 84.73    & 92.15        & 89.47    && 85.47    & 94.53 & 90.08    && 99.93    & 99.58    & 99.88    &&        2.2    & 3.3    & 4.1\\
		\midrule 
		25     quantile        &77.88    &87.94    &74.29        && 72.82    & 88.65&73.26    &&99.82    &99.15    &99.70    &&        1.4    &2.0    &2.0  \\
		\midrule 
		75     quantile        & 90.21    & 94.81    & 93.98    && 91.97    & 97.28 & 95.16    &&99.98    &99.83    &99.97    &&    4.1    &5.1    &10.3\\

		%
		
		
		
		\midrule[0.1pt]
	\end{tabular}

	\end{table}

		\begin{table}[!t]
		\centering
		\caption{DSCs and HD95 of the synthetic method on BraTS'19 Training set.} 
		\label{tab:segmentation_result_training}
		
		\begin{tabular}{@{\extracolsep{\fill}}c|cccc|cccc|cccc|ccc}
			\toprule[1pt] 
			\rule[-1ex]{0pt}{2.5ex}
			\multirow{3}{*}{\textbf{}} & \multicolumn{3}{c}{\textbf{Dice}} && \multicolumn{3}{c}{\textbf{Sensitivity}  }  && \multicolumn{3}{c}{\textbf{Specificity}  } && \multicolumn{3}{c}{\textbf{HD95 (mm)}  } \\
			\cmidrule[\heavyrulewidth]{2-4} 
			\cmidrule[\heavyrulewidth]{6-8}
			\cmidrule[\heavyrulewidth]{10-12}
			\cmidrule[\heavyrulewidth]{14-16}
			&\textbf{ET} &\textbf{WT} & \textbf{TC} && \textbf{ET} &\textbf{WT} & \textbf{TC}&& \textbf{ET} &\textbf{WT} & \textbf{TC}&& \textbf{ET} &\textbf{WT} & \textbf{TC}\\
			
\midrule 
Mean 				&79.26		& 91.65		& 90.76	&& 84.49	& 91.89 & 90.76	&& 99.86	&99.51	&99.77	&& 3.5	& 5.7	& 3.4 \\
\midrule 						
Std.      			&23.96		& 05.59		& 7.13	&& 14.46	& 08.04	& 08.17	&& 0.178	&0.47	&0.34	&& 7.3	& 11.0	& 4.6\\
\midrule 						
Median         		&87.04		& 93.29		& 92.88	&& 88.12	& 94.35	& 93.22	&& 99.92	&99.64	&99.88	&& 1.4	& 2.8	& 2.0\\
\midrule 
25     quantile     &79.49		& 89.89		& 88.34	&& 80.69	& 88.99	& 87.96	&& 99.831	&99.37	&99.74	&& 1.0  & 1.8	& 1.4\\
\midrule 
75     quantile   	&91.54		& 95.39		& 95.28	&& 93.78	& 97.23	& 96.43	&& 99.975	&99.80	&99.95	&& 2.2	& 4.9	& 3.6\\

			\midrule[0.1pt]
		\end{tabular}
	\end{table}

\section{Overall Survival Prediction Model}

BraTS'19 dataset contains 102 gross total resections (GTR) pre-operative scans out of 335 training cases in which the age of patients is available. These subjects are applied for developing a model to predict the overall survival (OS) of the patient. To this end, we measure the volume of WT, TC, and ET after segmentation to create a feature vector to predict patient OS. We also consider patient's age as an input feature to increase survival prediction accuracy. Thus, we have a 4-dimensional normalized feature vector that scaled between 0 and 1. We train different regression models to predict OS through supervised machine learning, including linear models, regression trees, support vector machines (SVMs) with different kernel functions, Gaussian process regression (GPR) models, and ensembles of trees. We measure root mean square error (RMSE), maximum absolute error (MAE), and prediction speed during inference  (observation/sec) to assess model performance. 
The 5-fold cross-validation is applied to evaluate these models with four feature vectors.

Table \ref{tab:linear}	presents linear regression models, including linear, interactions, robust, and stepwise linear models. We also evaluate regression Trees with three minimum leaf sizes, i.e., 4, 12, and 36 in this table.

\begin{table}[!t]
	\centering
	\caption{Comparison between linear models and regression trees with different hyperparameters.} 
	\label{tab:linear}
	\begin{tabular}{@{\extracolsep{\fill}}c|cccc|cccc}
		\toprule[1pt] 
		\rule[-1ex]{0pt}{2.5ex}
		\multirow{3}{*}{\textbf{}} & \multicolumn{4}{c}{\textbf{Linear Regression Models}} && \multicolumn{3}{c}{\textbf{Regression Trees}}   \\
		\cmidrule[\heavyrulewidth]{2-5} 
		\cmidrule[\heavyrulewidth]{7-9}
	
		&\textbf{Linear} &\textbf{Interactions} & \textbf{Roubust} & \textbf{Stepwise} &&\textbf{Fine} & \textbf{Medium}& \textbf{Coarse} \\
		
		\midrule 
		RMSE 			& 316.81 	& 375.23 & 326.76	& 314.07	&& 377.46 & 317.35	& 327.95 \\
		\midrule 						
		MAE      		& 224.24	& 250.04 & 220.04	& 223.36	&& 277.04	& 237.8	& 237.38	\\
		\midrule 						
		Pred. speed    	& 2000		& 6200	 & 7800	    & 7600	    && 4900	& 19000	& 19000\\
				
		\midrule[0.1pt]
	\end{tabular}
\end{table}

Table \ref{tab:svm}	evaluates SVMs models through different Kernel functions and scales.  We consider kernel scales 0.5, 2, and 8 for fine, medium, and coarse Gaussian SVM, respectively.

\begin{table}[!t]
	\centering
	\caption{Comparison between different SVM kernels. Kernel scales for Gaussian (Gaus.) SVM are considered as 0.5, 2, and 8 for Fine, Medium, and Coarse, respectively.} 
	\label{tab:svm}
	\begin{tabular}{@{\extracolsep{\fill}}c|c|c|c|c|c|c}
		\toprule[1pt] 
		\rule[-1ex]{0pt}{2.5ex}
		\multirow{3}{*}{\textbf{}} & \multicolumn{6}{c}{\textbf{SVM}}  \\
		\cmidrule[\heavyrulewidth]{2-7}

		&\textbf{Linear} &\textbf{Quadratic} & \textbf{Cubic} & \textbf{Fine Gaus.} &\textbf{Medium Gaus.} & \textbf{Coarse Gaus.} \\
		
		\midrule 
		RMSE 			& 323.92 	& 354.44 & 377.65	& 349.41	& 341.52 & 329.36	 \\
		\midrule 						
		MAE      		& 220.02	& 244.46 & 263.68	& 234.66	& 228.45	& 221.86		\\
		\midrule 						
		Pred. speed    	& 5400		& 16000	 & 17000    & 16000	    & 17000	& 15000	\\
		
		\midrule[0.1pt]
	\end{tabular}
\end{table}

Table \ref{tab:ensemble} shows GPR and Ensemble Trees models. The former is evaluated with squared exponential, Matern 5/2, exponential, and rational quadratic kernel functions. The boosted Trees and the Bagged Trees are examined for the latter.

\begin{table}[!t]
	\centering
	\caption{Comparison between GPR and ensemble models with several kernel functions. The abbreviation is: (Exp)onential} 
	\label{tab:ensemble}
	\begin{tabular}{@{\extracolsep{\fill}}c|c|c|c|c|cc|c}
		\toprule[1pt] 
		\rule[-1ex]{0pt}{2.5ex}
		\multirow{3}{*}{\textbf{}} & \multicolumn{4}{c}{\textbf{Gaussian Process Regression Models}} && \multicolumn{2}{c}{\textbf{Ensemble Trees}}   \\
		\cmidrule[\heavyrulewidth]{2-5} 
		\cmidrule[\heavyrulewidth]{7-8}
		
		&\textbf{Squared Exp.} &\textbf{Matern} & \textbf{Exp.} & \textbf{Rational Quadratic} &&\textbf{Boosted}&\textbf{Bagged} \\
		
		\midrule 
		RMSE 			& 332.28 	& 344.9 & 344.2	  & 332.28	&& 344.16 & 333.36	 \\
		\midrule 						
		MAE      		& 237.93	& 250.37 & 249.95 & 237.93	&& 251.42	& 240.69		\\
		\midrule 						
		Pred. speed    	& 4900		& 12000	 & 13000  & 13000	    && 2600	& 3400	\\
		
		\midrule[0.1pt]
	\end{tabular}
\end{table}

\begin{figure*}[ht]	
	\includegraphics[width=12cm]{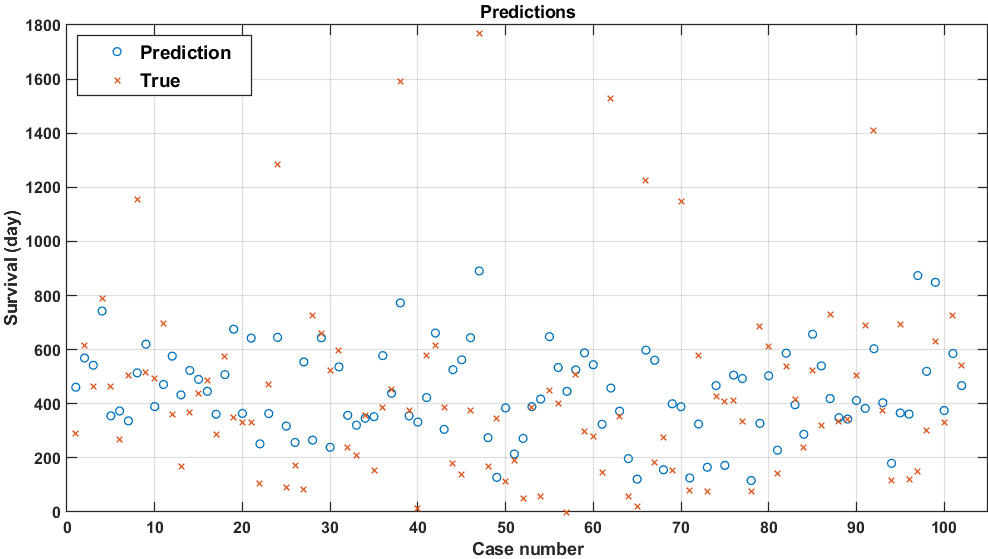}
	\caption{Survival prediction per day through the stepwise linear regression model. The predicted results versus case number.}
	\label{fig:os_4features}
\end{figure*}

Fig. \ref{fig:os_4features} displays predicted response versus subject numbers in BraTS'19. The predictions are accomplished with the stepwise linear regression model.

\begin{figure*}[ht]	
	\includegraphics[width=12cm]{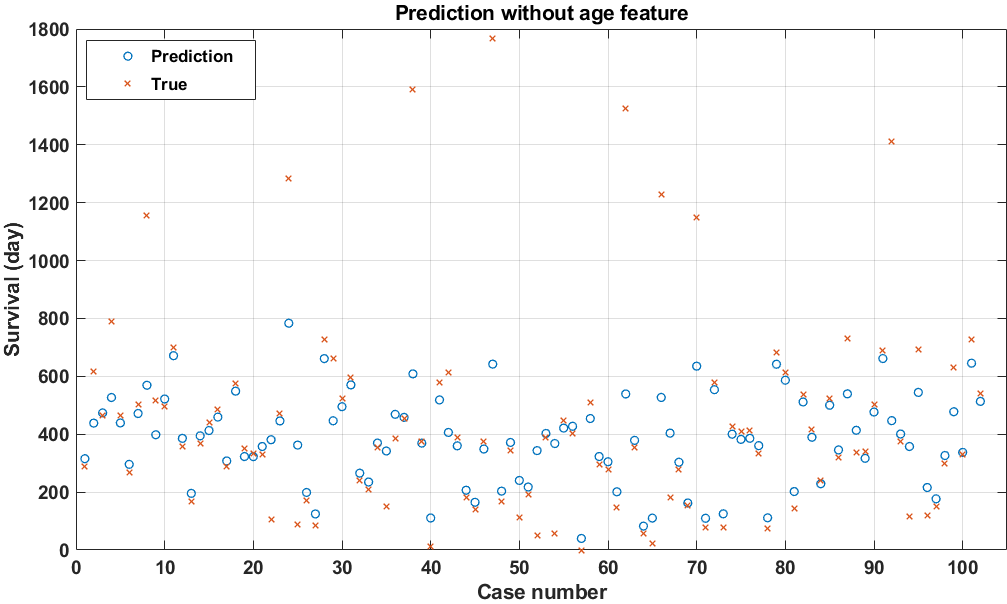}
	\caption{Survival prediction per day through the stepwise linear regression model. The predicted results versus case number.}
	\label{fig:os_3features}
\end{figure*}

Fig. \ref{fig:os_3features} also illustrates predicted response based on three features. We removed age feature to evaluate the effect of this feature on OS task. Table \ref{tab:age} compare RMSE with and without age feature for survival task. 

\begin{table}[!t]
	\centering
	\caption{RMSE with and without age feature.} 
	\label{tab:age}
	\begin{tabular}{@{\extracolsep{\fill}}c|c|c|c|c|c}
		\toprule[1pt] 
		\rule[-1ex]{0pt}{2.5ex}
		
		\textbf{Feature numbers}&\textbf{Linear} &\textbf{Regression Trees} & \textbf{SVM} & \textbf{Ensemble} &\textbf{GPR} \\
		\midrule[0.1pt]
		
		RMSE with age feature    & 314.07 & 317.35 & 323.92 & 333.36 & 332.26 \\
		\midrule 
		RMSE without age feature & 357.96 & 361.45 & 351.99 & 362.62 & 352.53 \\
		
		\midrule[0.1pt]
	\end{tabular}
\end{table}

	\section{Conclusion}
	\label{sec:Conclusion}
	
This paper provided a framework for the synthetic segmentation that translated FLAIR MR images into high-contrast synthetic MR ones for segmentation.  Synthesizing based on the GAN network empowers our model to decrease the number of real channels in multimodal brain tumor segmentation challenge 2019. We also implemented several regression models to predict the OS of each patient. We found that the stepwise linear model overwhelmed other traditional regression models in terms of RMSE. We also observed that patient age as a distinctive feature in the OS prediction tasks.

	\section{Acknowledgment}
	\label{sec:Acknowledgment}
	This work was supported partly by National Natural Science Foundation of China (Nos.61871274, 61801305, and 81571758), National Natural Science Foundation of Guangdong Province (No. 2017A030313377), Guangdong Pearl River Talents Plan (2016ZT06S220), Shenzhen Peacock Plan (Nos. KQTD2016053112 051497 and KQTD2015033016 104926), and Shenzhen Key Basic Research Project (Nos. JCYJ20170413152804728, JCYJ20180507184647636, JCYJ20170818142347 251, and JCYJ20170818094109846).

	%
	%
	%
	%
	\bibliographystyle{splncs04}
	\bibliography{mybib_brats}

\end{document}